\documentclass[twocolumn]{aastex631}

\usepackage{graphicx}	
\usepackage{amsmath}	
\usepackage{amssymb}	

\usepackage{mathtools}
\usepackage[normalem]{ulem}

\definecolor{darkgreen}{rgb}{0.13, 0.55, 0.13}
\definecolor{rubinered}{rgb}{0.82, 0.0, 0.34}
\definecolor{orange(ryb)}{rgb}{0.98, 0.4, 0.01}
\definecolor{scut}{rgb}{0.58, 0.42, 0.11}
\newcommand{\nwmcut}[1]{}

\newcommand{\cdmcut}[1]{}

\newcommand{\timespan}{50 Myr}

\newcommand{\code}[1]{\texttt{#1}}
\newcommand{\epsff}{\epsilon_{\rm ff}}
\newcommand{\Msun}{\rm M_\odot}
\newcommand{\rvec}{{\mathbf{r}}}
\newcommand{\vvec}{{\mathbf{v}}}

\begin{document}

\title[Playing with FIRE]{Playing with FIRE: A Galactic Feedback-Halting Experiment Challenges Star Formation Rate Theories}

\author[0000-0003-1053-1262]{Shivan Khullar}
\affiliation{David A. Dunlap Department of Astronomy \& Astrophysics, University of Toronto, 50 St. George St., Toronto, ON M5S 3H4, Canada}
\affiliation{Canadian Institute for Theoretical Astrophysics, University of Toronto, 60 St. George St., Toronto, ON M5S 3H8, Canada}

\author[0000-0001-9732-2281]{Christopher D. Matzner}
\affiliation{David A. Dunlap Department of Astronomy \& Astrophysics, University of Toronto, 50 St. George St., Toronto, ON M5S 3H4, Canada}

\author[0000-0002-8659-3729]{Norman Murray}
\affiliation{Canadian Institute for Theoretical Astrophysics, University of Toronto, 60 St. George St., Toronto, ON M5S 3H8, Canada}

\author[0000-0002-1655-5604]{Michael Y. Grudi{\'c}}
\affiliation{Carnegie Observatories, 813 Santa Barbara St, Pasadena, CA 91101, USA}

\author[0000-0001-5541-3150]{D{\'a}vid Guszejnov}
\affiliation{Center for Astrophysics, Harvard \& Smithsonian, 60 Garden Street, Cambridge, MA 02138, USA}

\author[0000-0003-0603-8942]{Andrew Wetzel}
\affiliation{Department of Physics \& Astronomy, University of California, Davis, Davis, CA 95616, USA}

\author[0000-0003-3729-1684]{Philip F. Hopkins}
\affiliation{TAPIR, Mailcode 350-17, California Institute of Technology, Pasadena, CA 91125, USA}




\begin{abstract}

Stellar feedback influences the star formation rate (SFR) and the interstellar medium of galaxies in ways that are difficult to quantify numerically, because feedback is an essential ingredient of realistic simulations. To overcome this, we conduct a feedback-halting experiment starting with a Milky Way-mass galaxy in the FIRE-2 simulation framework. Terminating feedback, and comparing to a simulation in which feedback is maintained, we monitor how the runs diverge. We find that without feedback, interstellar turbulent velocities decay. There is a marked increase of dense material, while the SFR increases by over an order of magnitude. Importantly, this SFR boost is a factor of $\sim$15-20 larger than is accounted for by the increased free fall rate caused by higher densities.  This implies that feedback moderates the star formation efficiency per free-fall time more directly than simply through the density distribution.  To probe changes at the scale of giant molecular clouds (GMCs), we identify GMCs using density and virial parameter thresholds, tracking clouds as the galaxy evolves. Halting feedback stimulates rapid changes, including a proliferation of new bound clouds, a decrease of turbulent support in loosely-bound clouds, an overall increase in cloud densities, and a surge of internal star formation. Computing the cloud-integrated SFR using several theories of turbulence regulation, we show that these theories underpredict the surge in SFR by at least a factor of three. We conclude that galactic star formation is essentially feedback-regulated on scales that include GMCs, and that stellar feedback affects GMCs in multiple ways.

\end{abstract}



\section{Introduction} \label{sec:Intro}

The energetic and kinematic 
feedback from young stars impacts the evolution and interstellar properties of galaxies by cycling matter from dense to diffuse or from cold to hotter phases.  Feedback 
-- which encompasses supernovae (SNe), massive-star winds, radiation pressure, photo-ionization, and protostellar winds, among other effects -- has been implicated in  clearing the environments of OB associations, in stirring and destroying giant molecular clouds (GMCs) and their substructures, in expelling matter that would otherwise collapse into stars or star clusters,  in driving the motions that limit gravitational instability within galactic disks, and in launching matter into the circum-galactic medium. Despite general consensus on these points, important questions remain to be settled.

One of these concerns the role of stellar feedback in the star formation rate (SFR), 
for which we focus on two classes of theory. 
\textit{Feedback-regulated} theories (as in \citealt{ThompsonQuataertMurray05} and \citealt{OstrikerShetty11}) 
hold that  stars form as quickly as needed in order for their feedback to support a gaseous environment 
against its own weight, or to maintain the (mostly turbulent) kinetic energy required to stave off gravitational instability.   \textit{Turbulence-regulated} theories \citep{Klessen00Collapse,KM05,HC11,PN11} relate star formation to the rate at which  turbulent, self-gravitating clouds produce regions of localized collapse.  A common  feature of these theories is that $\epsff$, the SFR normalized by the free-fall rate (see e.g. \autoref{eq:def-epsff}), depends (to varying degrees) on the Mach numbers and virial ratios of the regions within which stars form -- that is, GMCs and dense molecular clumps within them. Critically, stellar feedback is not explicitly accounted for within these turbulence-regulated models.

It is important to note here that some turbulence-regulated theories address the dynamics of  gravitational collapse. For example, \citet{guszejnov2015a, guszejnov2016} use an excursion-set approach \citep{hopkins2012, hopkins2013, bond1991} to follow the hierarchy of collapsing sub-regions within GMCs. 
There is another class of theory, presented initially in \citet{zamora-aviles2012, zamora-aviles2014a} and consolidated in \citet{vazquez-semadeni2019}. These authors posit a global hierarchical collapse scenario where molecular clouds are undergoing constant pressureless collapse rather than being near-equilibrium structures. They propose scaling relations between the SFR and the mass of clouds derived while taking into account mass inflow and ionization from massive stars.

Feedback-regulated and turbulence-regulated models are not mutually exclusive:
after all, feedback drives turbulence, as do shear, collapse, and accretion \citep[e.g.][]{forbes2023}.  
To clarify, our working definition of a turbulence-regulated model is one in which the SFR is predicted by a few dimensionless parameters of turbulence and is  not directly related to the driving mechanism of the turbulence or feedback in any other way. These theories attempt to capture all the effects of stellar feedback, and other physical processes through changes in these dimensionless parameters. In feedback regulated models,  by contrast, the SFR is determined by the interplay between the strength of feedback and environmental parameters like the weight of the ISM \citep[see e.g.][]{KimOstriker11_DiskRegulation, kim2013, OstrikerKim2022, Sun23_SFRin80gals}.

Both feedback-regulated and turbulence regulated models enjoy considerable observational support. Feedback regulation explains many properties of disk galaxies, including the Kennicutt-Schmidt relation \citep{ThompsonQuataertMurray05, OstrikerShetty11}, in which the SFR per unit area is a power law of the gas mass per unit area. Turbulence regulation is consistent with the low and reasonably constant values of  $\epsff$  inferred within individual GMCs \citep{pokhrel2021} and in regions of star cluster formation \citep{KrumholzTan2007}. Feedback regulated models can also predict low values of $\epsff$ if a small fraction of the ISM is in gravitationally bound structures \citep{OstrikerKim2022}. Turbulence regulation can explain the galactic star formation relations if the galactic SFR is just the sum of the small contribution per GMC. 
On the other hand, in some cases, turbulence regulated models can be in contention with observations. Observed values of $\epsff$ are around 1\% (with a scatter of $\sim$0.3-1 dex). However, the multi-freefall versions of the turbulence regulated models \citep{FK12}, which are extensions to the original theories, can predict values of $\epsff \sim 100\%$ or greater in certain situations (highly compressive turbulence with high Mach numbers, $\mathcal{M}>10-20$).


Additionally, significant variations in $\epsff$ in the Galactic center \citep{Kruijssen04_GalctrSF}, within nearby galaxies \citep{Sun23_SFRin80gals}, and on GMC \citep{Lee2016,Ochsendorf2017} and sub-GMC  \citep{Wells22_SFR_within_clumps} scales all suggest that turbulence regulation should be examined carefully. 
From a theoretical perspective, too, the importance of stellar feedback even in the formation of individual stars \citep{Matzner00_efficiency, F15}  questions the notion of a clean separation  between the scales of turbulence and feedback regulation.  One goal of this work, therefore, is to conduct an experiment within which feedback and turbulence regulation can be carefully disentangled. 

Another open question, related to what controls the SFR, involves how strongly stellar feedback affects the properties and populations of GMCs.   The consequences of stellar feedback processes are expected, at least in theory, to destroy, erode, and stir up the clouds and their contents \citep{Whitworth79,WilliamsMcKee97,matzner2002,murray2011,geen2016, HowardPudritz17_destruction,kim2018, Benincasa2020, menon2023}. Modern support for the strength of these feedback effects is found in the scale-dependent decorrelation between dense gas tracers such as CO and star formation tracers such as H$\alpha$ or free-free radio emission \citep{schruba2010, kruijssen2014, kruijssen2018, chevance2020, semenov2021,mcleod2021, kim2022a}. 
However, some simulation studies conclude that feedback is too weak to affect GMCs in this way \citep{dobbs2015,tasker2011,tasker2015} .
If so, this would imply that stellar feedback can only influence GMCs through its impact on the diffuse medium in which they form (presumably via  supernova explosions). However, these authors did not account for photo-ionization, stellar winds, or radiation pressure, all of which are expected to be important for GMC destruction \citep{Krumholz2009, Fall2010, hopkins2012b, 2010ApJ...709..191M,Chevance2019a}. Therefore, a second goal for our work is to provide a  clear quantitative reevaluation of the effectiveness of stellar feedback on GMCs, within a realistic galactic context.  

\begin{figure*}[ht]
    \centering
    \includegraphics[width=\textwidth]{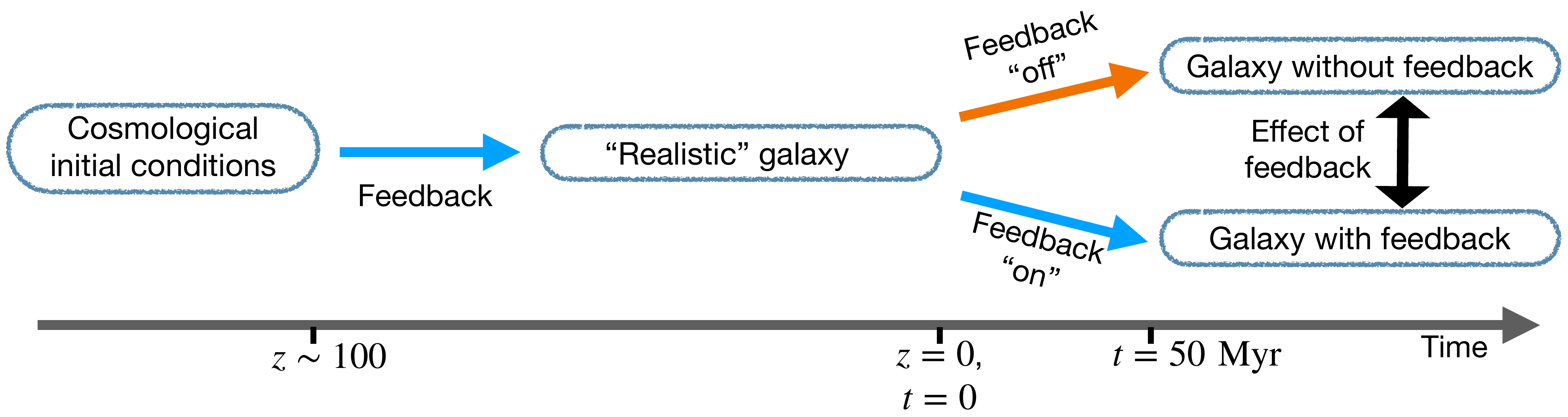}
    \vspace*{-15pt}
    \caption{A schematic describing our numerical experiment set up. We use a FIRE-2 galaxy evolved from cosmological initial conditions and create two branches in its evolution: one where we keep stellar feedback active and the other where we turn it off. This allows us to isolate the effect of feedback on the galaxy.}
    \label{fig:enter-label}
\end{figure*}

Somewhat ironically, the fact that  stellar feedback is central to galactic evolution poses an obstacle when one wants to quantify its effects in a realistic simulation, simply because feedback is needed to make a simulation realistic.  This makes it difficult to compare its ability to erode GMCs to those of tidal gravity and cloud collisions, for instance, or to evaluate its effect on $\epsff$ in a galaxy like the Milky Way.

To overcome this problem we conduct  controlled {\em feedback-halting} experiments using the Feedback In Realistic Environments (FIRE-2, \citet{FIRE-2Hopkins2018}) simulations, also used, for instance, by \citet{Benincasa2020}. Over cosmic time (up to $z=0$) our galaxy experiences
multiple realistic feedback effects: photo-ionization, photoelectric
heating, stellar winds, radiation pressure, and time-resolved SN explosions (protostellar outflows are not included).   At time $t=0$ our galaxy experiences two hypothetical futures: one in which feedback is maintained, and another in which it is cut off.  (In fact we run multiple scenarios, testing subsets of feedback phenomena, but we focus primarily on the all-or-none distinction in this paper.)  We then examine how the two scenarios  diverge over the subsequent \timespan, paying special attention to the influence of feedback on the properties of GMCs and the rate of star formation within them.  Using identical initial conditions makes our comparisons relatively insensitive to the randomness introduced by stochastic events; and because the shared initial state reflects a sophisticated array of feedback effects acting over cosmic time, it is as realistic as we can hope to achieve in the current state of the art.   

For our analysis we will inspect the overall distribution of interstellar properties within the galaxy, as well as the properties of GMCs that we identify and track. 
Because we focus on galactic scales, our resolution is poor compared to high resolution simulations of individual GMCs
On the other hand, we have the advantage of thousands of GMCs in our cloud samples with varying initial configurations and galactic environments, which allows us to compare their evolution statistically with self-consistent initial conditions.
Among other comparisons, we study the impact of stellar feedback on the star formation laws obeyed by these GMCs and on the distribution of star formation efficiency defined on the cloud scale. 

We detail our simulations, cloud identification and tracking methods in \autoref{sec:Methods}. In \autoref{sec:Results} we discuss the impact of halting feedback on the interstellar properties of the galaxy, the global SFR, and the character of GMCs.  In \autoref{S:Theory-Comp} we make a detailed comparison to several turbulence-regulated theories for the SFR.  We discuss our results in the context of previous work, and provide caveats, in \autoref{sec:Discussion}. We distill our conclusions in \autoref{sec:Conc}.

\begin{figure*}
    \centering
    \includegraphics[width=\textwidth]{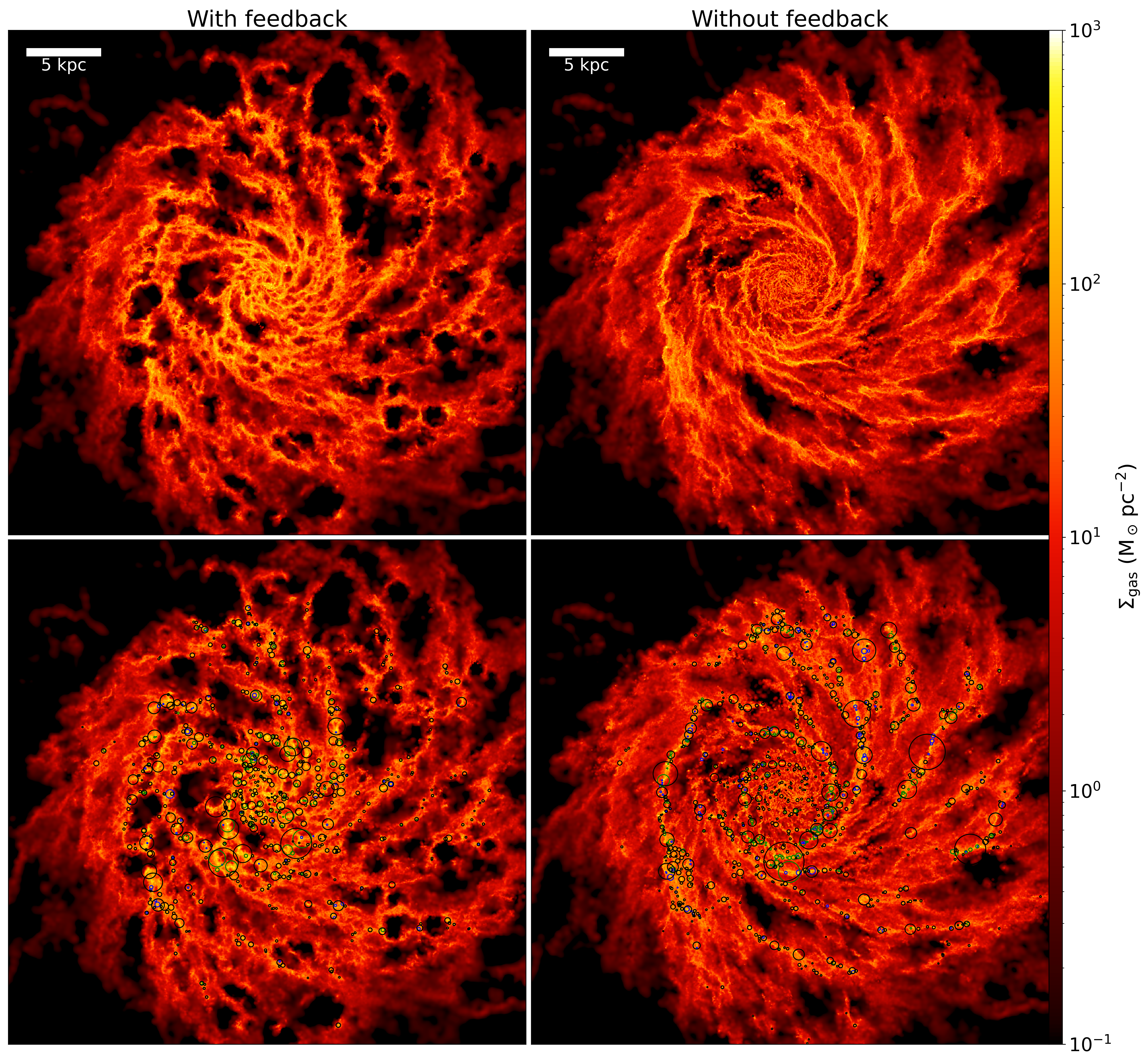}
    \vspace*{-20pt}
    \caption{
    Gas column density in the control (left panels) and feedback-halting simulation (right panels) after \timespan\ of divergent evolution.  On the lower panel, 
   circles display molecular clouds identified with various  selection criteria.  Blue, green, and black circles denote n10v2, n10v5, and n10v10 clouds (naming indicates \code{n\_min},  \code{alpha\_crit}). 
    Each cloud is represented by a circle of corresponding to its effective radius. 
    }
    \label{fig:MW-GMC-vir-comp}
\end{figure*}

\section{Simulations and Methods}
\label{sec:Methods}

We use a Milky-Way like galaxy (\textit{m12i}) from the Feedback in Realistic Environments (FIRE-2) suite of simulations \citep{Wetzel2016,FIRE-2Hopkins2018}. 
These evolve primordial perturbations using the {\sc{music}} code \citep{Hahn2011} up to redshift $z=100$. Using these initial conditions, the simulation is evolved up to $z=0$ in a large cosmological box using the Meshless-Finite Mass (MFM) hydrodynamics code {\sc{GIZMO}} \citep{Hopkins2015} at low resolution. These simulations are then re-run at higher resolution in a sub-region around the galaxy under study.  To track the physics of the interstellar medium, they include metallicity-dependent heating and cooling across a temperature range of 10-10$^{10}$\,K in addition to explicitly modelling stellar feedback in the form of OB/AGB star winds, radiation pressure, photo-ionization, photoelectric heating, {and mechanical} feedback from type Ia and type II supernovae. Star formation (the conversion of fluid elements into star particles) occurs when gas cells satisfy all of the following criteria: {they are} (a) Jeans unstable, (b) self-gravitating, with a virial parameter less than unity, (c) self-shielding or molecular (as assessed using the \citealt{Krumholz2011} prescription), and (d) sufficiently dense: ($n >  n_{\rm crit}= 1000$ cm$^{-3}$). 
We refer the reader to Appendix C, D and E, \citet{FIRE-2Hopkins2018} for more details on the star formation and stellar feedback methods used in this work.

\begin{table*}
    \centering
    \begin{tabular}{|c|c|c|}
        \hline \hline
             \textbf{Symbol} & \textbf{Name} & \textbf{Calculated as} \\
         \hline \hline
         M$_{\rm gas, tot}$ & Total gas mass in clouds & \\
        M$_{\rm gas}$ & Instantaneous cloud mass& \\
              R$_{\rm eff}$ & Instantaneous effective radius & $\sqrt{5/3 \langle (\rvec - \rvec_{\rm cm})^2\rangle_M}$\\
        A$_{\rm proj}$ & Area of the projected (face-on galaxy view) cloud's convex hull & \\
        R$_{\rm gal}$ & Instantaneous galactocentric radius & \\
                $\delta M_*$ & Protostellar Mass of cloud within R$_{\rm eff}$ ($\delta t_*$=1 Myr) & \\
        M$_{\rm *, tot}$ & Total stellar mass of tracked cloud over its lifetime & \\
                t$_{\rm ff, mult}$ & Instantaneous (multi) free-fall time & $\langle ({32 G \rho / 3\pi})^{1/2}\rangle_M^{-1}$\\ 
           SFR & Star formation rate & $\delta {\rm M}_{*}/\delta {\rm t_{*}}$ \\
        SFR$_{\rm tot}$ & Total star formation rate in all clouds & \\
                $\epsff$ & Star formation efficiency per free-fall time & ${\rm SFR}\times {\rm t}_{\rm ff, mult}/{\rm M_{\rm gas}} $ \\
        SFE & Star formation efficiency & ${\rm \delta M}_{\rm *}/({\rm M_{\rm gas}} + {\rm \delta M}_{\rm *})$\\
        SFE$_{\rm int} (t)$ & Integrated star formation efficiency & ${\rm M}_{\rm *, tot}(t)/({\rm M_{\rm gas}} + {\rm M}_{\rm *, tot}(t))$ \\
        $\sigma_{\rm v}$ & Velocity dispersion & 
             $\langle\frac13(\vvec-\vvec_{\rm cm})^2\rangle_M^{1/2}$ \\
                     $T$ & Temperature & $\langle T\rangle${$_M$} \\
        $c_s$ & Thermal sound speed   & ${\langle k_B T/(\mu m_H) \rangle}_M^{1/2}$ \\ 
        $\mathcal{M}$ & Mach number & $\sigma_{v}/c_s$ \\
        $\alpha_{\rm vir}$ & Virial parameter & $2(E_{\rm kin}+E_{\rm thermal})/E_{\rm grav}$ \\ 
        \hline \hline
    \end{tabular}
    \caption{Definitions of cloud properties used in this work. The subscripts $i$ and `cm' used above denote gas particles in the cloud and the cloud's centre of mass respectively. Here $\langle\cdots\rangle_M$ indicates a mass-weighted average within the cloud (and across cloud population if accompanied by another subscript), $\langle\cdots\rangle_{\rm clouds}$ denotes the average of a quantity across the cloud population. For tracked clouds (see text for definition), we use $\langle\cdots\rangle_t$ to denote the temporal mean of a quantity from Column 1.}
    \label{tab:Quants}
\end{table*}

{We use a snapshot of the galaxy at $z$=0, which we define as $t$=0, and create two primary branches {in its evolution,} 
one in which we keep the original physics and one in which we turn off all stellar feedback. We then evolve these two simulations up to $t$=\timespan \ (recording a snapshot every 1 Myr) and compare them to isolate the impact of feedback on the evolution of the GMCs. At $z=0$, the scale height of the galaxy increases with galactocentric radius from 
$\approx0.4$ to $1.5$ kpc \citep{gurvich2020}. 
Its virial radius  is $R_{\rm vir}$ = 275\,kpc and the virial mass is $M_{\rm halo}^{\rm vir}=1.2\times 10^{12} M_{\odot}$.  
{The total mass of the galaxy within a vertical extent $h_{\rm scale}=400$\,pc and radial extent R$_{\rm gal} = 25$\,kpc 
is $3.7\times 10^{10}$ M$_{\odot}$, 
of which gas comprises $0.9\times 10^{10}$ M$_{\odot}$.  The finest mass resolution for gas elements is $\sim$7000\,$M_\odot$. }

To define and detect the GMCs within the galaxy we use \code{CloudPhinder}, a method based on the SUBFIND algorithm \citep{Springel2001}, which was originally used to identify the largest self-gravitating bound structures that are present in the simulation \citep{Guszejnov2020}. CloudPhinder uses two input parameters to identify these structures: (i) \code{nmin}, the density cut-off and (ii) \code{alpha\_crit}, the minimum value of the virial parameter: 
\begin{equation}
    \alpha_{\rm vir}
    = \frac{2(E_{\rm kin}+E_{\rm thermal})}{|E_{\rm grav}|}
\end{equation}
in which $E_{\rm kin}$, $E_{\rm thermal}$, and $E_{\rm grav}$ are the bulk kinetic energy 
{(i.e., turbulent and rotational energies, in the centre-of-mass frame)}, thermal energy, and {self-}gravitational potential 
energy {of the cloud gas}, respectively.  
Starting at each density peak and examining successively lower density regions, the algorithm groups fluid elements into a cloud one at a time, accepting each one if the resulting group satisfies the density and virial parameter thresholds; see Appendix A of \citet{Guszejnov2020} for more details.  We limit our analysis to clouds composed of at least 32 fluid elements (this is based on the particle kernel size), setting a lower mass limit of $10^{5.35} M_\odot$. Typical massive GMCs in our simulations consist of a few hundred fluid elements. 

We consider three different input parameter combinations, {each of which defines its own cloud population:} (i) `n10v2' (\code{nmin}=10 cm$^{-3}$ and \code{alpha\_crit}=2), (ii) `n10v5', and (iii) `n10v10', all defined similarly.
We also add the suffix 'nofb' to specify the outcome of the run in which feedback is shut off. Because of the proliferation of options, we choose  `n10v5' (with and without feedback) to define our default cloud population. 


{In general, lower values of \code{nmin}, or higher values of \code{alpha\_crit}, cause more matter to be included within a cloud (sometimes combining separate clouds into one).  Therefore each cloud defined in `n10v2' is embedded within a cloud defined in any of the other combinations.}

In \autoref{fig:MW-GMC-vir-comp}, we show 
the distribution of gas surface density for the feedback-on (left panels) and feedback-off (right panels) simulations after \timespan\ of divergent evolution.  
The lack of supernovae feedback leads to a lack of bubbles in the galaxy without feedback, especially in the centre of the galaxy. 
In the bottom panels we overlay the GMCs identified using different input parameters to \code{CloudPhinder}, with the size of the circles denoting their effective radius R$_{\rm eff}$ (see \autoref{tab:Quants} for definitions of $R_{\rm eff}$ and other quantities). 
Raising \code{alpha\_crit} (or lowering \code{nmin}) results in fewer, larger clouds. 

Once we have a catalog of GMCs for each snapshot, we link them across time using {our own} algorithm, \code{CloudTracker}\footnote{\hyperlink{https://github.com/shivankhullar/CloudTracker}{https://github.com/shivankhullar/CloudTracker}} \citep{khullarCloudTracker}. \code{CloudTracker} tracks the particles in our GMCs using the identities of individual fluid elements. 
{Starting from $t=0$ and sorting clouds in a descending order by mass, we identify descendants of the original cloud population. If a cloud X has more than one child, we choose the child with the most mass donated from the parent. If this child is already a descendant of a more massive parent Y, we choose the heir next in line as the descendant of cloud X. If two clouds undergo a merger, we consider the less massive cloud to be dead.} 
Because most of our analysis does not require us to track clouds, we reserve the term 
`tracked cloud' for a GMC we track using \code{CloudTracker}; 
a `cloud' or `GMC' simply means an object identified by \code{CloudPhinder}.

\section{Results}
\label{sec:Results}

\begin{table}
    \centering
    \begin{tabular}{|c|c|c|c|c|} \hline \hline
        \textbf{Name} & \multicolumn{2}{c|}{\textbf{Untracked Clouds}} & \multicolumn{2}{c|}{\textbf{Tracked Clouds}}  \\ \hline
          & Mean &  \% SF & Total & \% SF \\ \hline \hline
         n10v2 & 127 & 19.7\% & 1824  & 42.8\% \\ 
         n10v5 & 379 & 21.6\% & 3606 & 31.8\%\\
         n10v10 & 479 & 20.1\% & 3787 & 25.7\%\\
         \hline
         n10v2 nofb & 450 & 60.1\% & 4538 & 65.1\% \\ 
         n10v5 nofb & 618 & 60.3\% & 5073 & 58.0\%\\
         n10v10 nofb & 632 & 60.4\% & 4804 & 56.5\%\\
         \hline \hline
    \end{tabular}
    \caption{Statistics on the GMC populations in our study after $t=$10 Myr (with feedback: top 3 rows, without feedback: bottom 3 rows).  Column 1: The input parameter combination for CloudPhinder (see \autoref{sec:Methods}). Column 2: The mean number of clouds identified per snapshot. Column 3: The mean percentage of star forming (SF) clouds per snapshot. Column 4: The total number of tracked clouds (starting from $t=10$ Myr) (see \autoref{sec:Methods} for the distinction). Column 5: The percentage of tracked clouds that form stars in their lifetime. The majority of GMCs do not form stars in their lifetime when feedback is on. Even without feedback, the percentage of star forming GMCs is about $\sim$60\%. With feedback, on average, only $\sim$20\% of clouds are star forming at any given time.}
    \label{table:stats}
\end{table}


\begin{figure*}
    \centering
    \includegraphics[width=\textwidth]{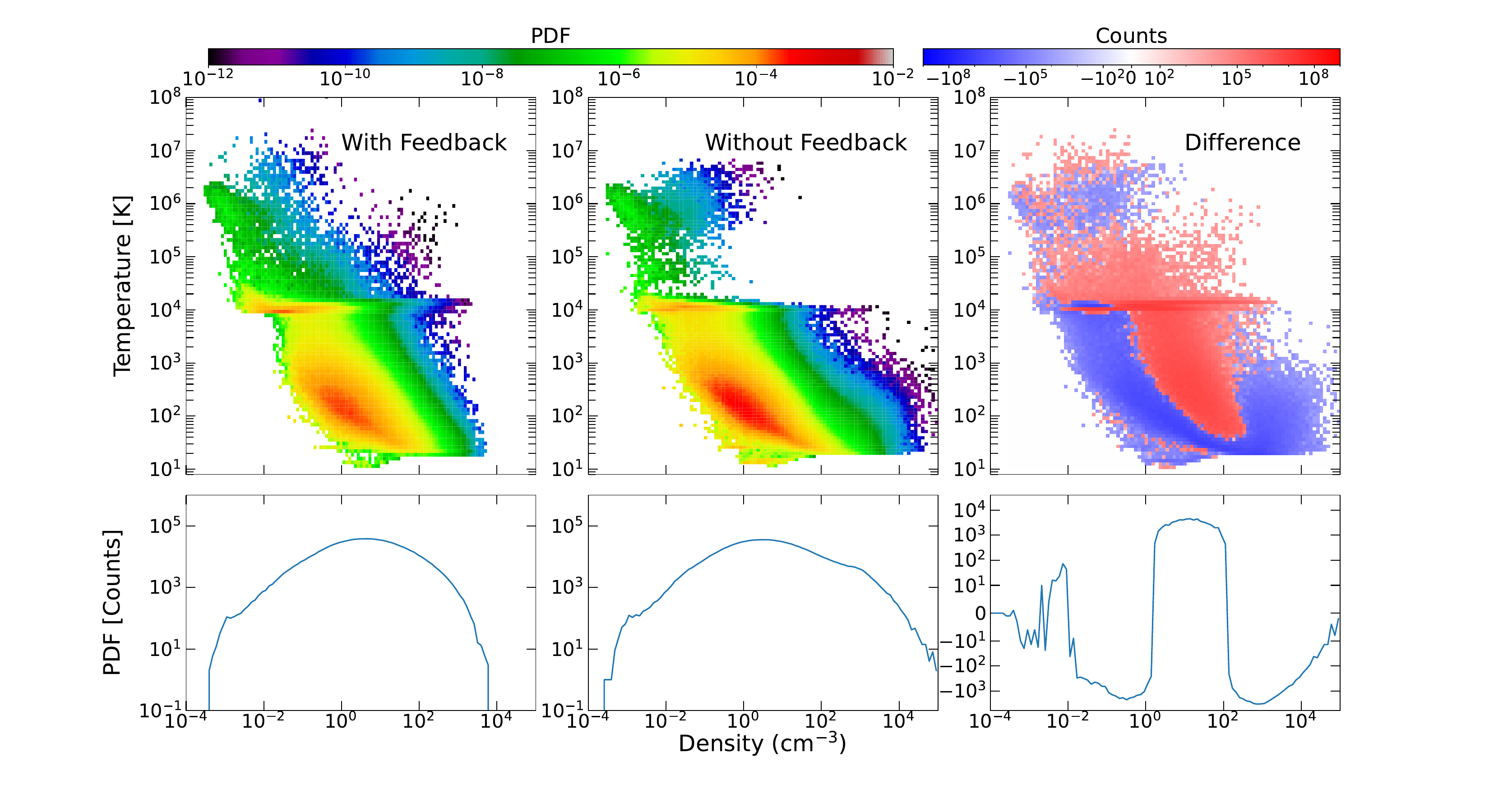}
    \vspace*{-30pt}
    \caption{
    \textit{Left:} A {mass-weighted} histogram of the temperature-density phase space for the galactic disk in the simulation with feedback, evaluated at $t$=\timespan. \textit{Centre:} Same, but for the simulation without feedback. \textit{Right:} The difference plot obtained by subtracting the centre panel's unnormalized (mass-weighted) value from the left panel. The red points show the gas cells present in the simulation with feedback but not in the one without. The blue points show the opposite. Gas flows from the red regions to the blue regions {after the termination of feedback.} 
    The bottom panels show just the density PDF of the gas in the respective simulations and again, the bottom-left panel shows the difference in the counts. The feedback-halting run develops an excess at both low and high densities. The increase in low density matter can be explained by the filling-in of voids created by feedback, whereas the increase in the cold dense matter can be explained by unopposed gravitational collapse.}
    \label{fig:Temp-dens}
\end{figure*}

\begin{figure}
    \centering
    \includegraphics[width=\columnwidth]{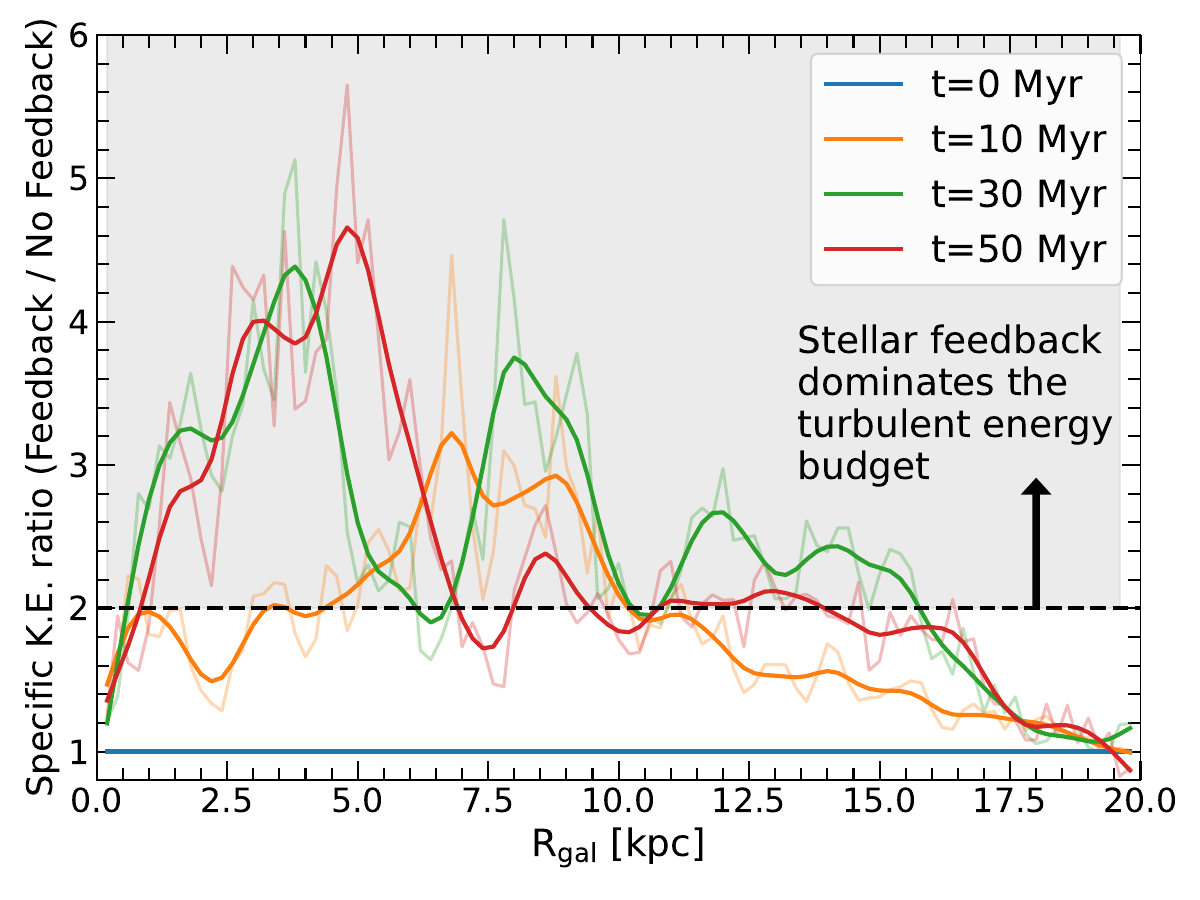}
    \vspace*{-20pt}
    \caption{Evolution of turbulent kinetic energy within the feedback-halting experiment, compared to the control simulation (solid lines represent cubic spline fits to the data). Halting feedback suppresses the velocity dispersion of the ISM. Stellar feedback is the dominant contributor to the turbulent energy budget compared to other sources of turbulence above the dashed line.}
    \label{fig:Turbulence}
\end{figure}

\begin{figure}
    \centering
    \includegraphics[width=\columnwidth]{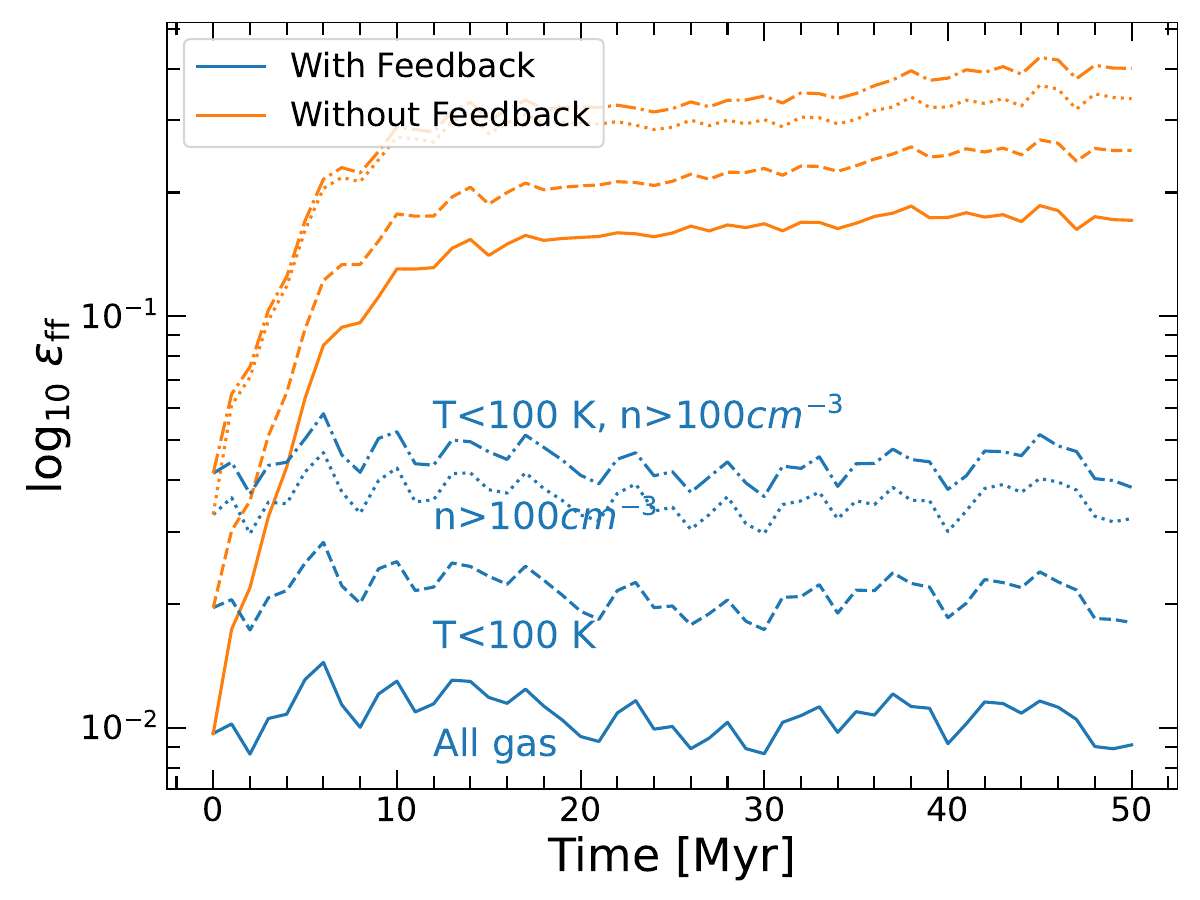}
    \vspace*{-20pt}
    \caption{
    Star formation efficiency per free-fall time $\epsff$, computed for the entire gaseous interstellar medium (solid line), and also for cold (dashed line), dense (dotted line), and dense \& cold (dash-dotted line) subcomponents. The no-feedback run displays a rapid surge in $\epsff$ by about one order of magnitude, relative to the control simulation, regardless of which portion of the gas we consider.}
    \label{fig:Epsff_galaxy}
\end{figure}

We begin by examining the overall impact of halting feedback on the galaxy's interstellar medium and star formation rate, starting with the distribution of matter in the phase space of density and temperature. 

In \autoref{fig:Temp-dens} we display the mass-weighted 2D histograms of $(n_H,T)$ and 1D density PDFs, as well as their differences, for a snapshot of the entire galactic disk (galactocentric radius $R<$25\,kpc,  altitude from the midplane $|z|<0.4$\,kpc) at $t=$\timespan.  

Examining the 2D histograms, we observe 
an accumulation of especially dense, cold matter in the no-feedback case -- leading to an increase at both the low- and high-density ends of the distribution (negative regions on the top right panel of \autoref{fig:Temp-dens}). 
We interpret the increase of low-density matter as the filling-in of voids created by feedback prior to $t=0$, and the increase of cold, dense matter as gravitational collapse that is unopposed by stellar feedback. Dense photoionized gas is absent since there are no more HII regions in the simulation without feedback.  Note the appearance of gas above our star formation density threshold ($n>n_{\rm crit}$), which can exist because of the other requirements for star formation.  

In each column of \autoref{fig:Temp-dens} the bottom panel is the density PDF. 
In this representation, we see that the run with feedback displays a reasonably log-normal density PDF.   Halting feedback leads to a broadening of this PDF, with what appears to be a power law tail extending to high densities (corresponding to the excess of cold, dense matter seen in the top panels) and a shifting of the rest of the PDF to lower densities. The broadening at the high density end is expected and seen in idealized simulations of collapse and star formation on smaller scales \citep[see e.g.][]{kritsuk2011density, FK13, khullar2021}.
Because the gas mass is almost identical between the two branches of evolution, the two histograms have nearly equal areas; therefore the difference histograms (right panels) average almost precisely to zero. Our results here corroborate earlier findings by \citet{hopkins2011}.

We compare the specific kinetic energy of the galaxies with and without feedback by calculating the velocity dispersions in $\sim 0.5\times0.4\times0.4$ kpc sized chunks at different galactocentric radius ($R_{\rm gal}$) in \autoref{fig:Turbulence}. 
We see that halting feedback suppresses the velocity dispersion of the interstellar medium. The difference is greatest where most of the star formation occurs, 1-5\,kpc from the galactic centre, and very little beyond 15\,kpc, where few stars form. We can also infer that the contribution from stellar feedback dominates the turbulent energy budget within $\sim$15\,kpc. Turbulence is driven by feedback wherever the ratio of specific kinetic energy is greater than 2 in \autoref{fig:Turbulence}. This comes from the following observation: if $\sigma_{\rm v, fb}^2 + \sigma_{\rm v, all \ else}^2 = \sigma_{\rm v, total}^2$, then $\sigma_{\rm v, fb}^2 > \sigma_{\rm v, all \ else}^2$, if $\sigma_{\rm v, total}^2/\sigma_{\rm v, total - fb}^2 > 2$. This represents a conservative lower limit because the increased importance of gravity will add to the total velocity dispersion in the galaxy without feedback.

\autoref{fig:Turbulence} suggests that stellar feedback is the major contributor to the turbulent energy budget in the inner parts of the galaxy. This corroborates earlier findings by \citet{green2010} who find a correlation between the star formation and velocity dispersions of a sample of galaxies in the nearby Universe. Perhaps not surprisingly, our results are also consistent with the findings of \citet{orr2020} and previous work using the FIRE simulations \citep[see e.g.][]{hopkins2013b, hayward2017}. 

Other processes, such as shear, accretion, inflow, etc. dominate the turbulent energy budget at the outskirts of the galaxy. This is consistent with the conclusions of \citet{forbes2023}, who find accretion through the disk to be an important contributor to turbulence at large galactic radii. 

\subsection{Impact on Galactic Star Formation Rate and Efficiency}

One particularly dramatic outcome of halting feedback is a rapid increase in SFR, relative to the control simulation, amounting to a factor of 23.2 by $t=$\timespan.  Importantly, this surge far exceeds the increase of  the mass-averaged free-fall rate $\langle t_{\rm ff}^{-1}\rangle_M$ implied by the change in the density PDF; this rate increases only by a factor of 1.23.  That is, star formation efficiency per free-fall time 
\begin{equation}\label{eq:def-epsff}
\epsff = \frac{\dot{M}_*}{\langle t_{\rm ff}^{-1}\rangle_M M}
\end{equation} 
(where $\dot M_*$ is the SFR within the region of interest,  $M$ is its gaseous mass, and $t_{\rm ff} = (3\pi/32G\rho)^{1/2}$ is the local free-fall time)
itself increases by a factor of $23.2/1.23=18.8$, relative to the control simulation, when we terminate feedback.  These numbers change according to the region over which $\epsff$ is evaluated, as is visible in \autoref{fig:Epsff_galaxy}, but the effect persists.  If we  restrict our attention to just the cold gas ($T<100$\,K) we find that $\epsff$ increases by a factor of 14; for dense gas ($n_H>100$\,cm$^{-3}$) the increase is a factor of 10.5, and it is also a factor of 10.4 for gas that is both cold and dense.  

Some turbulence-regulated star formation theories, such as \citet{KM05}, predict that $\epsff$ should be a slowly-varying function of other dimensionless parameters, such as the turbulent Mach number and virial parameter of the region under consideration.  In \autoref{S:Theory-Comp} we will examine the degree to which the change in $\epsff$ is compatible with these theories. 

Note, also, how rapidly star formation reacts to the absence of feedback: $\epsff$ appears to change within the first Myr after feedback is halted and saturates at its new, higher value after about 15\,Myr.  The speed of this transition strongly implies that feedback is crucial in limiting star formation within individual GMCs, in addition to any effect it may be having on the stability of the galactic disk or on the creation of new GMCs. 

These results suggest that halting stellar feedback should have a marked impact on the internal dynamics and star formation rates within  individual GMCs.  For this reason we turn now to examining the GMC population. 

\subsection{Impact on GMCs}

\begin{figure*}
	\includegraphics[width=\textwidth]{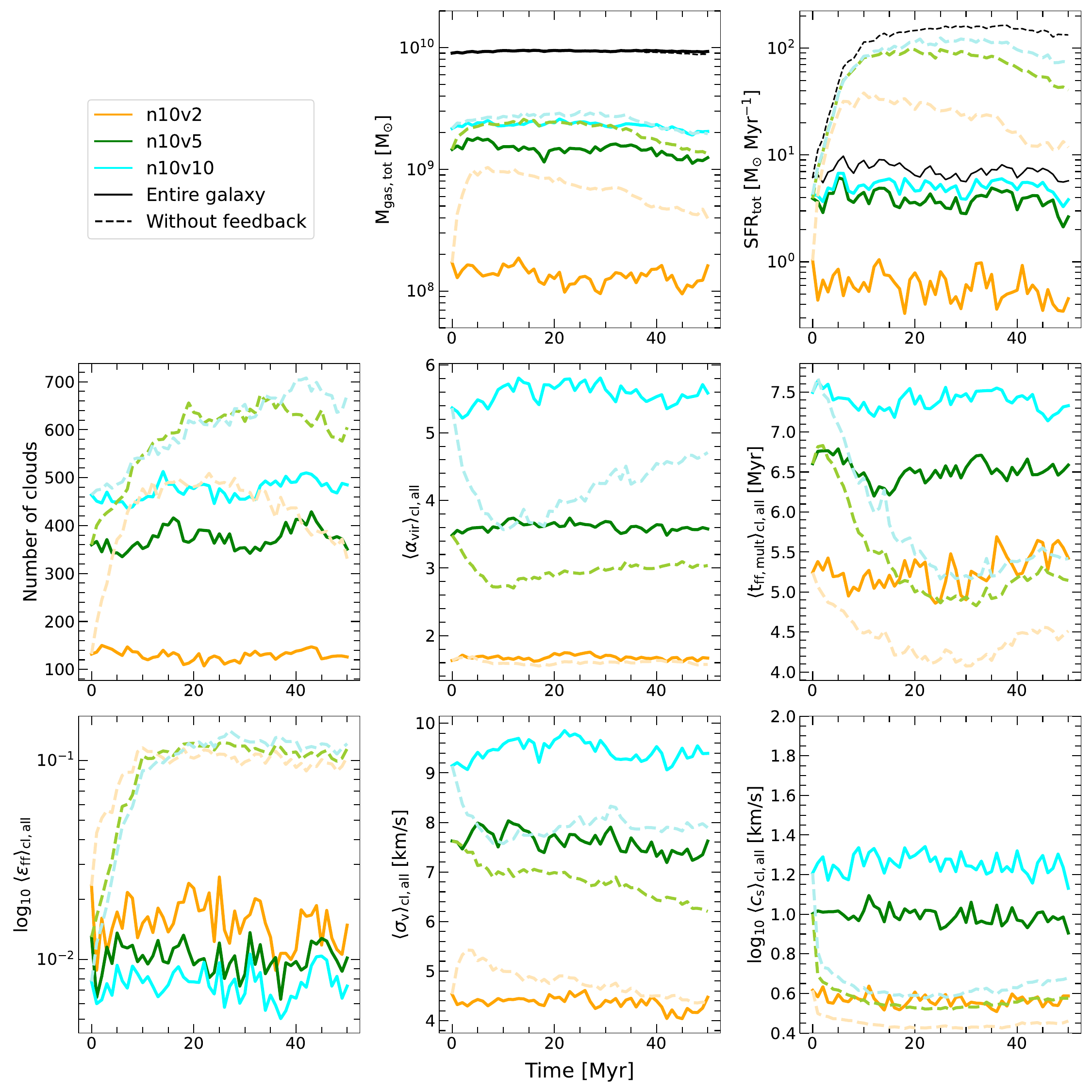}
    \vspace*{-20pt}
    \caption{Evolution of  cloud properties for all clouds identified within the galaxy.  Solid lines correspond to the clouds with a virial parameter cut-off value of 2, 5 and 10 respectively (see \autoref{sec:Results} for our naming convention). The dashed lines indicate the corresponding quantities from the simulation without feedback. All quantities are averaged over the entire cloud population at a given time. The black lines show the total gas mass and the total SFR in the galaxy with (solid) and without (dashed) feedback. The bottom-center and bottom-right panels show the gas mass and SFR in 3 different mass bins for the `n10v5' clouds. The different shades correspond to the different mass bins and light shades belong to the clouds without feedback. The thickness of these lines indicate the total number of clouds present at any given time in the galaxies. Stellar feedback keeps cloud masses and SFRs in check, maintains the number of clouds, their virial parameters and free-fall times. It also regulates the SFE per free-fall time at around $\sim 1\%$ overall. Note that the last two panels indicate an increase in the Mach number primarily due to a drop in the average sound speed.}
  \label{fig:OverallProp_evol}
\end{figure*}

In \autoref{table:stats} we summarize  important statistics of the cloud populations identified with various threshold parameters and for both the control and feedback-halting runs.  These define the sample for our cloud-level study of feedback effects.  We first consider the evolution of the cloud population as a whole, before turning to the effect of feedback on individual clouds tracked by \code{CloudTracker}.

In \autoref{fig:OverallProp_evol} we show the evolution of a number of properties of the GMC population, including the total mass and star formation rate within our various sub-populations (and in the galaxy as a whole) with and without feedback.  We concentrate on cloud samples with the same density threshold (\code{nmin}=10\,cm$^{-3}$) but various limiting virial parameters (\code{alpha\_crit}=2, 5, and 10).   

The effect of halting feedback within individual GMCs is very similar to what we saw for the galaxy as a whole: while the mass-averaged free-fall rate does rise, the SFR rises much more; so $\epsff$ increases by at least an order of magnitude.  
Comparing the feedback-halting run to the control simulation, we identify these additional trends: 
\vspace{-0.1in}
\begin{itemize} 
\item[-] Halting feedback does not alter the galactic gas mass greatly compared to the control simulation (\autoref{fig:OverallProp_evol}, top middle panel).  
\item[-] All our cloud categories (n10v2, n10v5, n10v10) gain mass when feedback is halted, by a similar amount (nearly $10^9\,M_\odot$: \autoref{fig:OverallProp_evol}, top center panel).  This is most significant for the bound clouds (n10v2), which comprise less than $10^8\,M_\odot$ at $t=0$.  These trends are mirrored by the growth in cloud numbers (middle left panel), however, the number of clouds still remains high long after turning off feedback.
\item[-] The surge in SFR stimulated by halting feedback (\autoref{fig:OverallProp_evol}, top right panel) is most significant for the bound clouds, but they display a surge in $\epsff$ (bottom left panel) comparable to the other cloud populations. It is worth noting that although our sub-grid prescription for star formation converts 100\% of the gas into stars on a local free-fall time, i.e. $\epsff\sim1$, even on marginally-resolved cloud scales the SFE is orders of magnitude smaller, $\epsff\sim0.01$. Naturally, this implies that GMCs be resolved with at least 100 cells.
\item[-] With feedback, the total SFR within less-bound clouds is significantly higher than within bound clouds  (top right panel), due to their higher overall mass, and the fact that star formation occurs in bound pockets but these may not always satisfy the minimum particle number restriction we impose during cloud identification.  When feedback is halted, the SFR surge in bound clouds makes up most of the difference -- implying that they become the sites of star formation within less-bound clouds. 
\item[-] Our least-bound clouds (n10v10) show a pronounced drop in virial parameter when feedback is halted, which is less pronounced in the intermediate population (n10v5) and almost absent in bound (n10v2) clouds (middle center panel). At the same time, the mean velocity dispersion drops in the unbound clouds, but rises in the most-bound clouds (bottom center panel). The free-fall times of all cloud populations decline relative to the control simulation (middle right panel); most significantly for the unbound clouds. 
\end{itemize} 
Although interpreting these trends is complicated by the appearance of density and virial parameter thresholds in our cloud definitions, we infer that when feedback stops, (1) high-$\alpha_{\rm vir}$ clouds lose turbulence, presumably both because of a loss in internal forcing and because of the declining velocity dispersion in the diffuse ISM; (2) collapse proceeds on all scales, increasing the number and velocity dispersion of the most-bound regions (but not their virial parameters, thanks to a selection effect from our cloud definition). We also note that the Mach numbers of our clouds tend to increase, on average, in the absence of feedback, due to the drop in the average sound speed (bottom right panel). Although not explicitly shown in this paper, many of the trends in our (time-averaged) GMC properties agree qualitatively with the differences found in the feedback and no-feedback galaxy-scale simulations of \citet{hopkins2012b}.

\subsubsection{Insights from  tracking clouds} \label{S:tracked-clouds} 

\begin{figure*}
	\includegraphics[width=\textwidth]{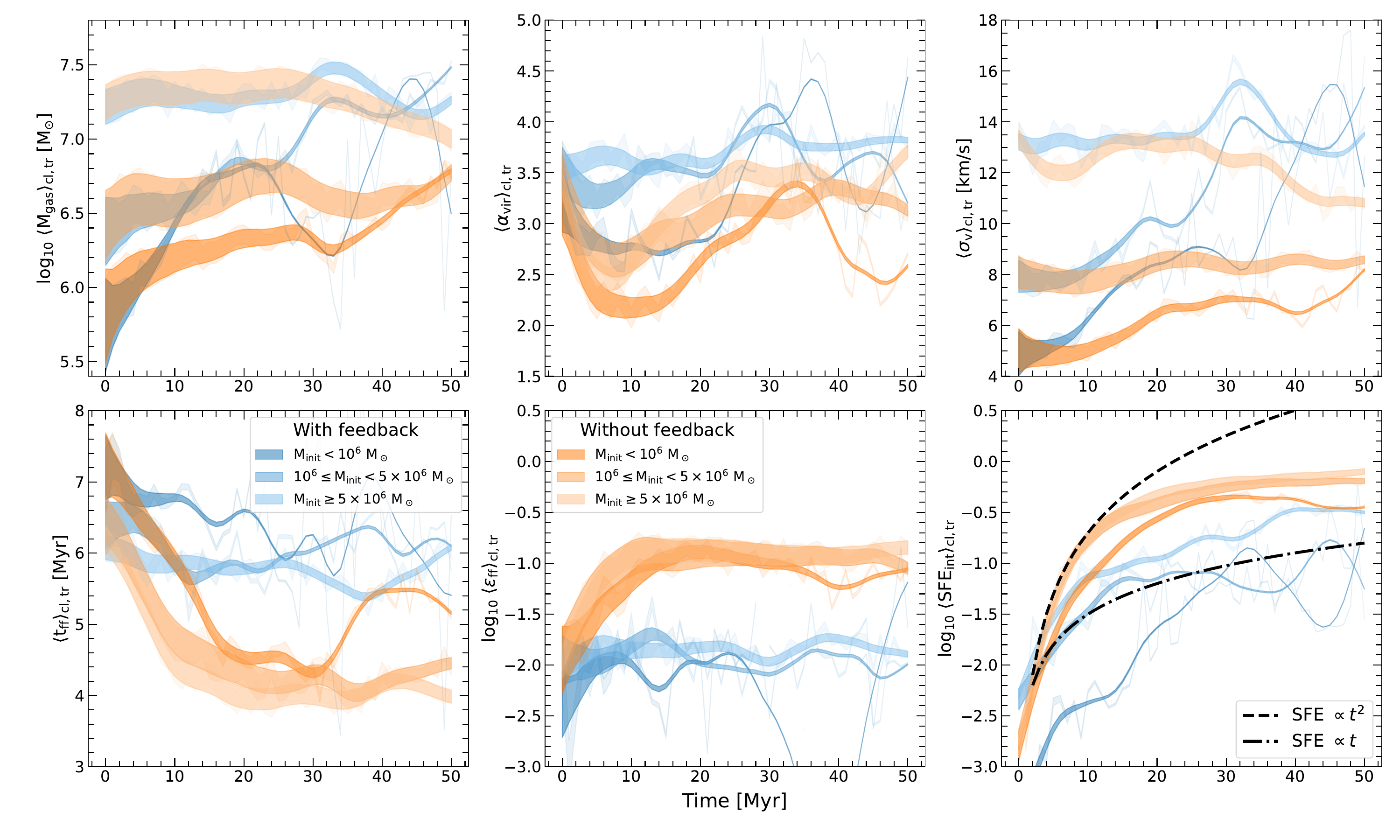}
    \vspace*{-20pt}
    \caption{We follow the evolution of different cloud properties (see \autoref{tab:Quants} for definitions) in 3 different initial mass bins for the `n10v5' clouds present at the start of our feedback-halting experiment. The blue lines represent clouds with feedback, whereas the  orange lines represent clouds without feedback. The quantities are averaged over the entire population at a given time. The thickness of the lines indicates the number of clouds averaged over, or the number of clouds still `alive' from $t=0$ Myr. Therefore, thin portions of the lines contain more variations since they are averaged over fewer surviving clouds.  Stellar feedback greatly impacts the evolution of the free-fall time, star formation efficiency per free-fall time, and total star formation efficiency. Feedback is also responsible for destroying low and intermediate mass clouds. Thus, although feedback seems to impact the virial parameter and velocity dispersion of surviving clouds minimally, survivor bias means that clouds that do not survive until late times had their virial parameters and velocity dispersions affected by feedback. Cloud properties are stable after $\sim$15 Myrs.} 
    \label{fig:Time_evol}
\end{figure*}

The appearance of additional clouds in our feedback-halting experiment complicates our view of the effect of feedback on GMCs.  For this reason we use \code{CloudTracker}, described in \autoref{sec:Methods}, to follow individual clouds. Here we only follow clouds present in the first snapshot (t=0).  (We note that our algorithm differs from that used by \cite{Benincasa2020}, but we forego a careful comparison with that work because we are not examining the detailed life histories of GMCs.) 

\begin{table*}
    \centering
    \begin{tabular}{|c|c|c|c|c|c|c|c|c|} \hline \hline
        \textbf{Time } & \multicolumn{4}{c|}{\textbf{With Feedback}} & \multicolumn{4}{c|}{\textbf{Without Feedback}}  \\ \hline 
           & Total &  Low mass & Intermediate mass & High mass & Total &  Low mass & Intermediate mass & High mass \\ \hline \hline
         0 & 360 & 171 & 125 & 64 & 360 & 171 & 125 & 64 \\
         10 & 103 (28.6\%) & 20 (11.6\%) & 42 (33.6\%) & 41 (64.1\%) & 203 (56.3\%) & 53 (31.0\%) & 89 (71.2\%) & 61 (95.3\%) \\
         20 & 47 (13.0\%) & 7 (4.1\%) & 13 (10.4\%) & 27 (42.2\%) & 160 (44.4\%) & 34 (19.8\%) & 69 (55.2\%) & 57 (89.1\%) \\
         30 & 33 (9.1\%) & 2 (1.2\%) & 9 (7.2\%) & 22 (34.4\%) & 130 (36.1\%) & 26 (15.2\%) & 53 (42.4\%) & 51 (79.7\%) \\
         40 & 28 (7.8\%) & 1 (0.6\%) & 7 (5.6\%) & 20 (31.2\%) & 95 (26.4\%) & 11 (6.4\%) & 41 (32.8\%) & 43 (67.2\%) \\
         50 & 19 (5.3\%) & 1 (0.6\%) & 3 (2.4\%) & 15 (23.4\%) & 72 (20.0\%) & 8 (4.7\%) & 30 (24.0\%) & 34 (53.1\%) \\
         \hline \hline
    \end{tabular}
    \caption{
    Summary statistics for the number of surviving clouds in different initial mass bins. Low mass: M$_{\rm init}<10^6 {\rm M_\odot}$, Intermediate mass: $10^6 {\rm M_\odot}\leq{\rm M}_{\rm init}<5\times10^6 {\rm M_\odot}$, High mass: $5\times10^6 {\rm M_\odot}\leq{\rm M}_{\rm init}$.
    Column 1: Time from the initial snapshot. Column 2: Total number of clouds surviving from the initial snapshot (in the simulation with feedback). Columns 3-5: Number of clouds surviving from the initial population in an initial mass bin 1, 2 and 3 respectively (in the simulation with feedback). Columns 6-9: Corresponding quantities for the simulation without feedback. Brackets denote percentages relative to the initial population. Feedback destroys low mass clouds more efficiently than high mass clouds.}
    \label{table:stats_lifetimes_table}
\end{table*}

\autoref{fig:Time_evol} shows how our tracked clouds evolve, focusing specifically on three ranges of initial mass (i.e., mass at $t=0$) within the n10v5 population.  Even more than before, it is important here to focus on the {\em difference} between the feedback-halting and control simulations, because defining the parent population $t=0$ causes the population to evolve even in the control simulation.  
We note that:
\vspace{-0.1in}
\begin{itemize} 
\item[-] $\epsff$ varies within these clouds much like it does in the galaxy as a whole (lower middle panel), with a burst of star formation of about an order of magnitude that is roughly independent of the initial mass; 
\item[-] While the stellar-to-gas mass fraction (SFE$_{\rm cl}$) grows roughly linearly in the control simulation, as one would expect for a constant cloud SFR, our feedback-halting experiment shows a period of SFE$_{\rm cl}\propto t^2$, indicative \citep{McKeeTan03hmsf,LeeChangMurray15sfr} of uncontrolled collapse and accretion (lower right panel).  This trend saturates at about 15\,Myr, when the stellar mass fraction is of order 30\%.  
\item[-] Stellar feedback is responsible for destroying low and intermediate mass clouds. The linewidths indicate the number of descendents of the original cloud population still surviving at a given time. We discuss this in more detail below.
\item[-] Tracked clouds' virial parameters drop significantly, relative to the control simulation, when we halt feedback -- although the difference is less clear for the lowest initial cloud masses (top middle panel). 
\item[-] Tracked clouds' free-fall times also drop precipitously, relative to the control population, when feedback ends (lower left panel).  
\end{itemize} 

These points, taken together, suggest that stellar feedback is an important regulator of the internal states of individual GMCs, and does not only act through its effect on the overall population of clouds.  Whether this is due primarily to momentum injection and turbulent forcing within the clouds themselves, as envisioned by \citet{McKee1989}, \cite{matzner2002}, and \cite{KrumholzMatznerMcKee06}, among others, or whether it is reflects a change in accretion \citep[e.g.,][]{klessenHennebelle10accretion,VasquezSemadeni10_Accretion,Goldbaum11} -- or another form of external driving --  remains to be determined. 

We summarize some statistics on the effects of feedback on cloud lifetimes in \autoref{table:stats_lifetimes_table}. The table lists the number of descendants of the original cloud population still surviving at a later time. The different initial cloud mass bins correspond to same intervals as \autoref{fig:Time_evol}. We adopt a loose definition for identifying cloud descendants as discussed in \ref{sec:Methods}. As a result, a significant population of high mass clouds exists in our sample compared to \citet{Benincasa2020} who find typical cloud lifetimes to be less than 20 Myr. Regardless, we find that stellar feedback is majorly responsible for destroying low and intermediate mass clouds. There are other mechanisms through which clouds can be disrupted, but a detailed comparison between the different mechanisms of cloud destruction is beyond the scope of this study. 

As a side note, we find that a significant fraction of our clouds are non-star-forming and remain that way for their entire lifespans (as seen in \autoref{table:stats}). This population is most pronounced at low masses and is more prevalent ($\sim$80\% vs 60\%) within the control simulation, relative to the feedback-halting run.  We speculate that these low-mass clouds are the result of turbulent compressions caused by stellar feedback, and that the absence of star formation is partly the consequence of our limited mass resolution. These low-mass non-star-forming clouds can also be part of a short-lived population which generally ends up merging into larger clouds before they form any stars.

\section{Comparison to Turbulence-Regulated SFR Theory} \label{S:Theory-Comp}

\begin{figure*}
	\includegraphics[width=\textwidth]{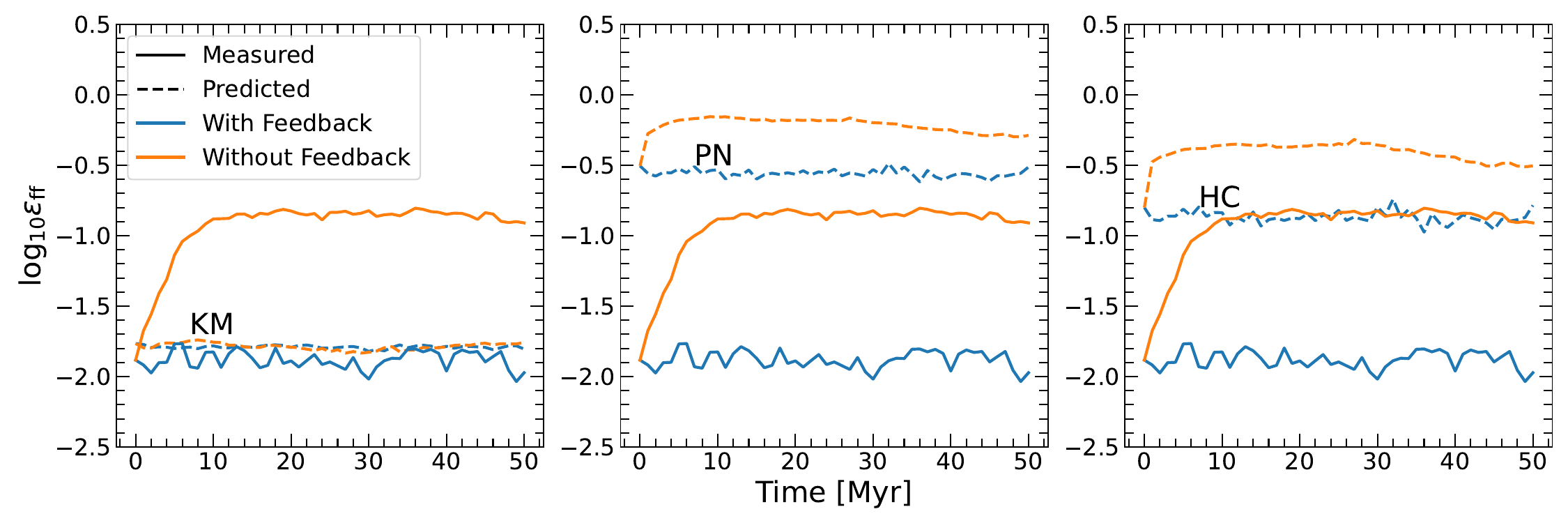}
    \vspace*{-20pt}
    \caption{Comparison between the population-averaged $\epsff$ within GMCs in our feedback-halting and control simulations, and the turbulence-regulated predictions of the KM, PN, and HC models. The turbulence regulated models differ both in terms of overall normalization  and in their predictions for the difference in $\epsff$ between the feedback and no feedback cases. None of the turbulence regulated models predict a change in $\epsff$ as large as we find in the simulation.}
    \label{fig:SFRcomparison}
\end{figure*}

\begin{figure*}
	\includegraphics[width=\textwidth]{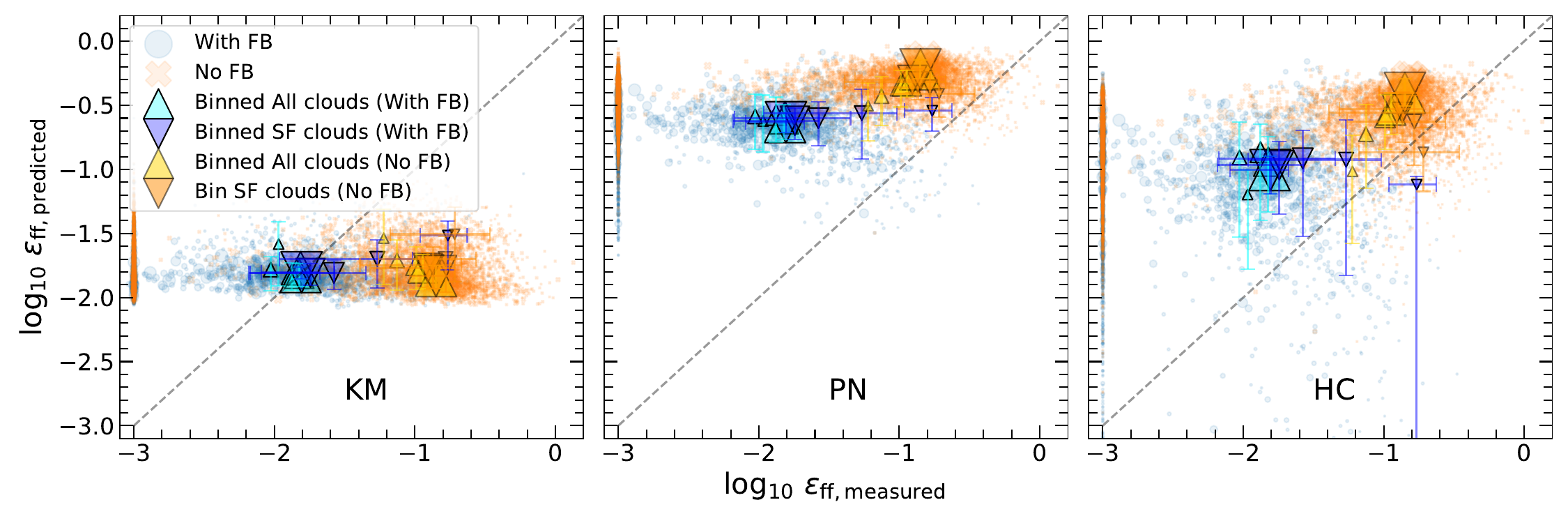}
    \vspace*{-20pt}
    \caption{Comparison between $\epsff$ within GMC in our feedback-halting and control simulations, and the turbulence-regulated predictions of the KM, PN, and HC models for individual tracked clouds (tracked $t$=10 Myr onwards). The blue circles and orange crosses indicate the mean $\epsff$ for tracked clouds with and without feedback respectively. Non-star forming clouds are indicated by the population at $\log_{10} \ \epsilon_{\rm ff,measured}=-3$. The upper triangles show the binned averages in 5 different mass bins along with 1$\sigma$ error bars for all clouds. The lower triangles only include the clouds which have non-zero star formation. The HC model comes the closest in predicting the right spread in $\epsff$, but it does so predominantly for the simulation without feedback. There is a trend for the simulated $\epsff$ in star forming clouds to decrease with increasing mass smaller downward pointing triangles have larger simulated $\epsff$ which the theories do not predict. This trend vanishes if all clouds are considered. }
    \label{fig:SFRcomparison-clouds}
\end{figure*}

Having identified GMCs, we are in a position to compare the surge in star formation we observe within our cloud population with predictions from the turbulence-regulated SFR theories of \citet{KM05}, \citet{PN11}, and \citet{HC11}  -- which we refer to as KM, PN, HC, respectively.  We do this by constructing the SFR of each GMC within one of these theories, calculating the net theoretical SFR, and comparing against the net SFR within our simulation.   Essentially, these theories adopt a log-normal (LN) form for the volumetric density probability distribution, so that the quantity $s = \ln \rho/\rho_0$ (for reference density $\rho_0$, usually taken to be the mean density) is normally distributed: 
\begin{equation} \label{eq:LN-PDF(rho)}
p(s)\propto \exp\left[-\frac{(s-s_0)^2}{2 \ln (1 + b^2 {\cal M}^2)}\right],
\end{equation} 
 where $b$ is a parameter that depends on how turbulence is driven, and lies in the range $\frac13\leq b\leq 1$ \citep{Federrath2010};  normalization sets $s_0$ and the coefficient of $p(s)$.   The star formation efficiency per free-fall time is then related to the portion of this PDF above a certain critical density threshold: 
\begin{equation}
\label{eq:epsff}
    \epsff = \frac{\epsilon}{\phi_t} \int_{s_{\rm crit}}^{\infty} \frac{t_{\rm ff}(\rho_0)}{t_{\rm ff}^*(\rho)} \frac{\rho}{\rho_0} p(s) ds,
\end{equation}
where the constants $\epsilon$ and $\phi_t$ relate to the core-to-star formation efficiency \citep{Matzner00_efficiency} and the timescale for star formation, $t_{\rm ff}^*(\rho)$ is the effective free-fall time assumed in the theory, and $s_{\rm crit}$ is defined below in eqns. (\ref{eqn: scrit KM}) to (\ref{eqn: scrit HC}); the remaining symbols carry their usual meanings. 

The models differ in two ways: first, KM and PN adopt a constant effective free-fall time evaluated at the mean density and critical density, $s_{\rm crit}$, respectively, whereas HC takes it to be the local free-fall time.  That is, $t_{\rm ff}^*(\rho) = [ t_{\rm ff}(\rho_0), t_{\rm ff}(\rho_{\rm crit}), t_{\rm ff}(\rho)]$ in [KM, PN, HC], respectively, where $\rho_{\rm crit} = \rho_0\exp{s_{\rm crit}}$.  Second, the critical density is different for each model but depends on the virial parameter $\alpha_{\rm vir}$ and the Mach number $\mathcal{M}$ as
\begin{equation}\label{eqn: scrit KM}
    s_{\rm crit, KM} = \ln  \left[ \frac{\pi^2}{5} \phi_x \alpha_{\rm vir} \mathcal{M}^2 \right],
\end{equation}
\begin{equation}\label{eqn: scrit PN}
    s_{\rm crit, PN} = \ln \left[ 0.067 \theta^{-2} \alpha_{\rm vir} \mathcal{M}^2 \right],
\end{equation}
and 
\begin{equation}\label{eqn: scrit HC}
    s_{\rm crit, HC} = \ln \left[ \frac{\pi^2}{15} (y_{\rm cut}^{-2} \alpha_{\rm vir} \mathcal{M}^{-2} + y_{\rm cut}^{-1} \alpha_{\rm vir}) \right],
\end{equation}
where $\phi_x$, $\theta$ and $y_{\rm cut}$ are parameters in the KM, PN and HC models respectively. These parameters are calibrated using simulations in \citet{FK12}, but we use the values provided by the original authors in our analysis, i.e., $\phi_x=1.12$, $\theta=0.35$ and $y_{\rm cut}=0.1$. Our conclusions are not altered if we choose \citet{FK12} values. Note that $\epsff$ values predicted from the KM and PN models differ due to evaluating the free-fall time differently even though the form of the critical density is the same.

We determine theoretical predictions for $\epsff$ for each cloud, using its virial parameter and internal turbulent Mach number. For the turbulence $b$ parameter, we use again the values the original authors chose, i.e., $b=1,0.5,0.5$ for KM, PN, and HC respectively. We fix $b$ for all clouds in our analysis to these values. We test whether $b$ changes significantly between our two cloud populations, with and without feedback, by stacking the density distribution of all clouds in each population and estimating $b$ from the width of the LN distribution generated (\autoref{eq:LN-PDF(rho)}). We find the difference in the two $b$ parameters to be close to zero. We note, however, that a more careful analysis measuring the $b$ parameter for each individual cloud could result in a differing conclusion.}


We compare the population-averaged $\epsff$ within our simulated clouds to these model predictions, for small and large values of the $b$ parameter, in \autoref{fig:SFRcomparison}.
First, we note that the KM model best reproduces the star formation rate of the control simulation, in which feedback remains operative -- that is, its default normalization is best in line with the numerical SFR as determined by FIRE's star formation prescriptions.  However,
the three theoretical predictions respond differently to the changes in the overall cloud properties: the HC model is more sensitive than PN, which is more sensitive than KM. None of the models predicts the magnitude of the SFR surge we observe; the most sensitive model (HC) showing only a factor of three change in $\epsff$ compared to a factor of 12 change  in the simulation.   

In \autoref{fig:SFRcomparison-clouds}, we show the predicted vs. measured $\epsff$ to check how well the models do for individual clouds. 
The scatter points show the time-averaged predicted and measured $\epsff$ for tracked clouds (tracked $t=10$ Myr onwards) in the feedback (blue circles) and no feedback (orange crosses) cases. There is a significant spread in the measured $\epsff$ ($\sim$ 0.5 dex) which the KM and PN models are unable to reproduce. The upward pointing triangles show the tracked star-forming cloud-average $\epsff$ in 5 different (evenly spaced in log) mass bins. Bigger triangles correspond to a higher mass bin. Downward-facing triangles are the same, but include non-star-forming clouds as well while computing the average. There is a clear trend of the measured $\epsff$ decreasing with increasing mass for star-forming clouds with feedback. This trend vanishes if we include non-star-forming clouds as well (most of which lie in the lowest mass bins).


At this stage, it is worth recapping the different assumptions between the theories. The critical density in the KM model is calculated as the density where the sonic scale is comparable to the Jeans length or, in other words, equating the gravitational potential energy of a Bonor-Ebert sphere to the energy in turbulent motions. PN derive the critical density by equating the mass of a Bonor-Ebert sphere to the mass of a uniform sphere with a radius equal to the thickness of the post-shock layer. They obtain a similar scaling with $\alpha_{\rm vir}$ and $\mathcal{M}$ (in the hydrodynamic limit) to the KM model, but determine that the relevant timescale is the free-fall time evaluated at the critical density of the cloud, instead of the mean density (as in KM). HC, by contrast, use a density dependent dynamical time and incorporate spatial information about the over-densities of various sizes in a cloud to calculate its SFR. For estimating the critical density, these authors assume the Jeans length at the critical density is equal to a fraction, $y_{\rm cut}$ of the cloud size. This fraction is set by largest fluctuations that can turn unstable to gravitational collapse in the formalism of \citet{hennebelle2008, hennebelle2009}. 

Our analysis also alludes to the fact that the correct timescale for star formation matters. The dependence on the cloud parameters ($\mathcal{M}$ in particular) increases as we account for the relevant timescale in an increasingly sophisticated manner (moving from KM to PN to HC). These theories assume a steady state scenario for star formation, or in other words, star formation occurs on timescales that are longer compared to the disruption of the cloud. Stellar feedback (especially early feedback), operates on shorter timescales and thus alters the evolution of the star forming region. 

Related to this, we also tested the multi-freefall versions of the KM, PN and HC models described in HC and \citet{FK12}. In the multi-freefall versions of KM and PN, the free-fall time is not evaluated at a particular density, and the dependence on density is preserved instead. Disregarding once again the issue of overall normalization, we find that the cloud population-averaged predicted $\epsff$ in the simulation without feedback increases by 2 orders of magnitude, far exceeding the single order of magnitude increase measured. The multi-freefall versions of the models are very sensitive to the $\mathcal{M}$ number. In our simulation without feedback, the temperature of the clouds is lower on average and as a result, the $\mathcal{M}$ increases. Despite the sensitive dependence on $\mathcal{M}$, the multi-freefall versions of the models do not reduce the scatter in $\epsff$ or match the measurements better. In fact, the multi-freefall KM and PN models predict values of $\epsff$ $\gg100\%$ for typical $\mathcal{M}$ numbers found in our cloud populations.


\section{Discussion}
\label{sec:Discussion}

The primary conclusions of our experiment are, first, that stellar feedback is critically important in limiting the rate of star formation on galactic and molecular cloud scales; and second, that simple theories for turbulence regulation fall short of predicting the surge in SFR that accompanies a loss of feedback.  In fact, there is a second problem:  the  PN and HC theories, which are more sensitive than KM to molecular cloud properties and therefore come closer to matching the SFR surge, tend to significantly over-predict the SFR in the presence of feedback (see \autoref{fig:SFRcomparison}) unless their normalization factors are adjusted to enforce it. 

Turbulence in the ISM can be driven by several different sources, one of which is stellar feedback. \textit{Of course the fact that turbulence regulates the rate of star formation is not in question}:  there is really no other way to describe why the mean SFR is so low.  Indeed, the effective {\em local} SFR within the simulation is \begin{equation}
\label{eq:epsff_galactic}
    \dot{\rho}_* = \left\{
        \begin{array}{ccl}
        \epsilon_{\rm ff,cell} {\rho_{\rm gas}}{t_{\rm ff}^{-1}},
            &~~&\mbox{where conditions met}  \\
          0, &~~&  \mbox{elsewhere}
        \end{array}
    \right.
\end{equation}
\citep[e.g.][]{semenov2019, hopkins2018} where  $\rho_{\rm gas}$ is the gas density, and the local conditions for star formation discussed in \autoref{sec:Methods} and \citet{FIRE-2Hopkins2018} are essentially a metric of the degree of turbulence on that scale.  
In our runs the local SFR prescription corresponds to $\epsilon_{\rm ff,cell}=1$. (To achieve this in practice, gas particles are individually converted to stars in a probabilistic fashion with the equivalent mean local cell-scale rate.)   Despite being highly efficient where it operates, when averaged over the scales of molecular clouds and larger, this procedure gives $\epsff\simeq 10^{-2}$ in equation (\ref{eq:def-epsff}) -- because of the inhibitory effect of turbulence.  \citet{agertz2013}, \citet{hopkins2011}, and \citet{hopkins2013a} show that the Kennicutt-Schmidt relation \citep{Schmdit1959, Kennicutt98} is recovered on galactic scales, regardless of the specific value of $\epsilon_{\rm ff,cell}$ adopted in \autoref{eq:epsff_galactic}, provided the feedback prescriptions are accurate, suggesting that feedback is the underlying mechanism in regulating star formation at these scales. 

\textit{The relevant question here is whether turbulence regulation can be disentangled from the influence of stellar feedback and boiled down to a predictive theory for $\epsff$ in terms of dimensionless parameters, as in the KM, PN, and HC models}. If these cannot reproduce our results, might a more elaborate model succeed?  One possibility, as mentioned in \autoref{S:Theory-Comp}, would be to modify the driving parameter $b$ to reflect the loss of feedback.  Note that $\epsff$ increases with $b$, all else being fixed -- so to match the SFR surge, $b$ would need to be higher without feedback than with it.  It is not clear to us if this is realistic.  We note that \citet{padoan2016} and \citet{pan2019}   find that SNe can drive solenoidal modes (implying that $b$ is intermediate between $1/3$ and $1$). These authors calculate $b$ from the velocity field, rather than the acceleration field to which the $b$ parameter more closely relates to in simulations. On the other hand, \citet{dhawalikar2022driving} find that shock-driven turbulence is strongly compressive (implying $b\simeq 1$).  In the latter case, there would be no room for $b$ to increase when feedback halts. 

Alternatively, one might appeal to other types of theories for star formation within self-gravitating turbulent clouds, such as theories that describe the onset of collapse in ways that build from the \citet{bond1991} excursion-set formalism \citep{hopkins2012,guszejnov2015a, guszejnov2016, guszejnov2016a}, or related theories for hierarchical collapse \citep[e.g.,][]{vazquez-semadeni2019}. One could also invoke theories for the development of a power-law tail in the density PDF \citep[e.g.][]{kritsuk2011density, Federrath2013, Burkhart2017, Burkhart2018, Chen2018, Jaupart2020, khullar2021}, a feature observed within GMCs \citep[][]{Schneider2013, Schneider2015, pokhrel2016, schneider2022}.  Or, one could consider theories for time-dependent star formation efficiency, as in \citet{murray2015}, \citet{LeeChangMurray15sfr},  and \citet{murray2017}.   The difficulty, however, is that none of these provides an explicit prediction for $\epsff$ that can easily be assessed in terms of local conditions such as dimensionless cloud properties.  


In \autoref{sec:Intro}, we argued for a distinction between feedback-regulated and turbulence-regulated models. Such a separation is possible if there is a clear distinction between the (large) scales on which feedback operates, and the (small) scales on which turbulence regulates the rate of collapse. On scales of entire clouds or larger, a feedback-regulated model \citep[e.g.][]{ostriker2011, grudic2018, grudic2019a, orr2019, ostriker2022} (on kpc scales) may be more appropriate for predicting star formation efficiencies. On the other hand, however, turbulence regulated models may not be appropriate to predict star formation efficiencies at these scales. The central idea behind the turbulence regulated models is that star formation efficiency depends only on  dimensionless parameters describing the turbulent environment. 
Our results lead us to question whether this is true, and whether it is actually possible to separate the effects of feedback from those of turbulence --  considering that feedback, when present, appears to impact the 
dynamics at all scales resolved within our simulation. In other words, it may not be possible to prescribe the star formation efficiency purely in terms of bulk turbulent properties, when feedback is operative. 
Thus, the turbulence-regulated models need to be ``feedback aware".

\begin{figure}
    \centering
    \includegraphics[width=\columnwidth]{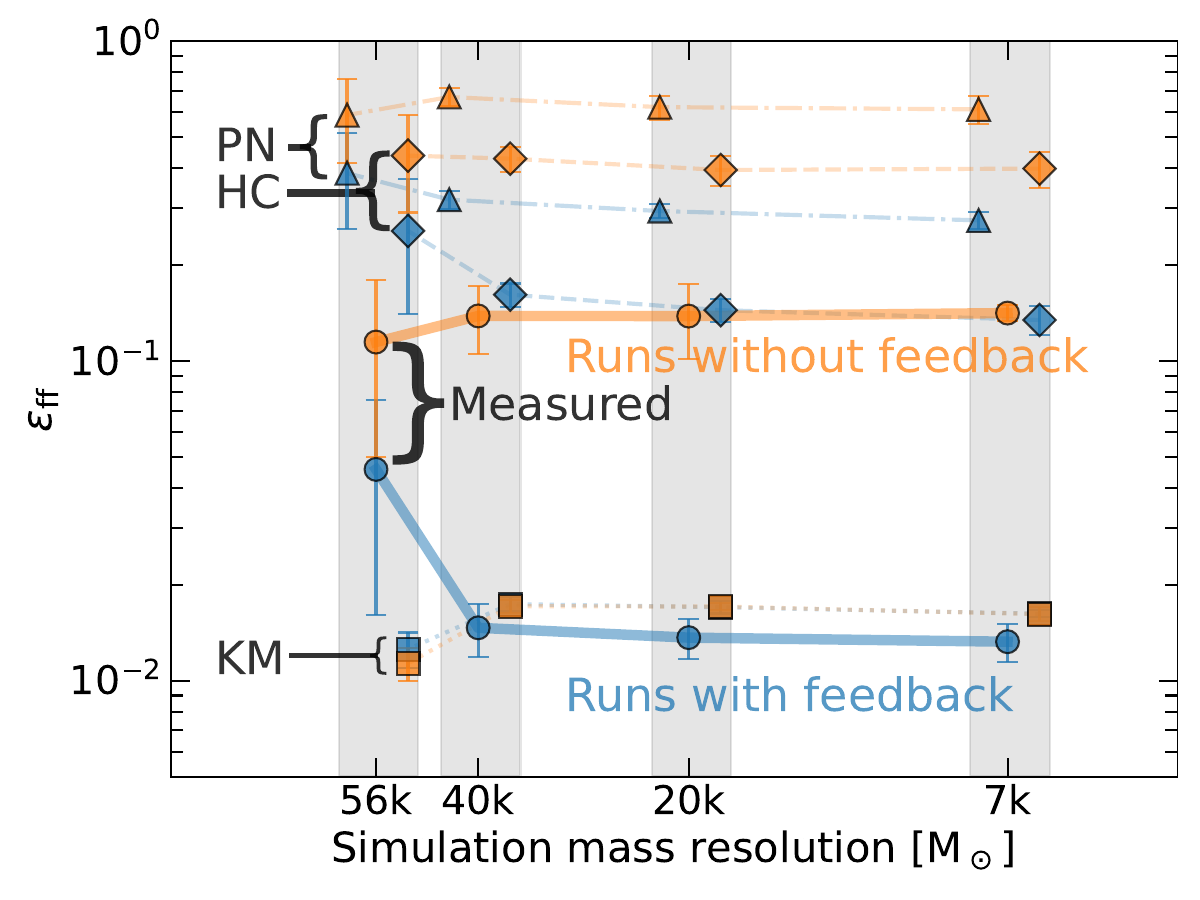}
    \vspace*{-15pt}
    \caption{The cloud population averaged $\epsff$ values (\autoref{fig:SFRcomparison}) for the feedback (blue) and no feedback (orange) cases, plotted as a function of simulation resolution (minimum particle mass). The predictions from the KM (squares), PN (triangles), and HC (diamonds) models are also plotted alongside the measurements (circles). Errorbars indicate the 1$\sigma$ temporal variation around the mean $\epsff$ after the first 10 Myr. The difference between the feedback and no feedback cases increases with increasing resolution (decreasing minimum mass) indicating that the effect is robust to resolution.}
    \label{fig:resolution}
\end{figure}

This work comes with one very important caveat:  at our  resolution we have only $\sim$ 140 fluid elements per $10^6\,M_\odot$, so a distribution like \autoref{eq:LN-PDF(rho)} cannot be fully resolved within any individual cloud.  
On the other hand, the PDF is well resolved for the cloud population as a whole. Considering that we see a consistent surge in $\epsff$ within individual clouds, the entire cloud population, and the overall ISM, we do not expect our conclusions to evaporate in future, better-resolved simulations. We show results from a resolution study in \autoref{fig:resolution}. We perform a resolution study by repeating our experiment at 4 different resolutions: 56,000 $\Msun$, 40,000 $\Msun$, 20,000 $\Msun$ and our fiducial 7,000 $\Msun$. The simulation at 56,000 $\Msun$ was run from $z~100$, in a similar manner as the 7,000 $\Msun$, branching out into two different evolutionary paths with and without feedback at $z$=0. However, the simulations at 20,000 and 40,000 $\Msun$ resolution were run by using a snapshot from the 7,000 $\Msun$ $\sim$ 240 Myr before $z$=0 and worsening resolution by merging particles. Once these simulations reach $z$=0, we stop particle merging and continue onwards creating two evolutionary branches for each simulation once again.

\autoref{fig:resolution} shows that the gap between the cloud population averaged $\epsff$ values in the feedback and no feedback cases tends to increase with increasing resolution (decreasing minimum mass). This indicates that the discrepancy between the turbulence models and the simulations increases with increasing resolution. We acknowledge that our results for convergence depend on the imprint an initially higher resolution galaxy leaves on $\epsff$ after merging particles. However, we defer a more careful comparison with higher resolution simulations, such as those of \cite{grudic2022}, to future work.

We also note that our results are limited to Milky-Way like galaxies, and may differ at high galactic column densities such as those found at high redshift. The importance of stellar feedback depends on the mean column density \citep[see e.g.][]{brucy2020, brucy2023}. While not explored in this work, we acknowledge that turbulence-regulated theories may work perfectly fine in these regimes without the need for any improvements.

\section{Summary and Conclusions}
\label{sec:Conc}

The fact that stellar feedback is critically important to galaxy evolution makes its effect hard to quantify without ruining the realism of a numerical simulation.  To evade this problem, we have conducted a controlled experiment in which we halt feedback within a normal spiral galaxy, evolved from cosmological initial conditions within the FIRE-2 simulation framework, and compare to a control simulation that is exactly the same at $t=0$ (after cosmic evolution) but in which feedback continues to operate.   Over the next \timespan\ the galaxy's interstellar medium reacts as one might expect to the loss of feedback: interstellar turbulence decays, holes from previous feedback start to fill in, and dense regions undergo an accelerated collapse -- along with a corresponding surge of star formation. We infer that stellar feedback is the dominant driver of turbulence at the $\sim 0.5$\,kpc scale, within $\sim 10$\,kpc of the galactic center. 

A remarkable aspect of this SFR surge is that it far exceeds what one would predict from the relative increase in the mass-averaged rate of free fall, even if the comparison is restricted only to dense and cold gas, or gas within clouds.  In other words, the star formation efficiency per free-fall time, $\epsff$ increases rapidly, by more than an order of magnitude, when feedback ends. This highlights the role stellar feedback plays in setting $\epsff$ \citep{suin2024}. 

Motivated by how quickly $\epsff$ increases -- it shows a significant change after 3-4\,Myr and saturates by 15\,Myr --  we examine the properties of molecular clouds identified using \code{CloudPhinder} with various virial parameter thresholds.  We see that feedback affects the populations and characteristics of less- and more-tightly bound clouds differently, but all clouds undergo a strong and rapid shift in their internal properties.   We find that the surge in $\epsff$ seen at the galactic level exists within GMCs as well. This is not simply the consequence of a changing cloud population, as we demonstrate by tracking individual clouds (using our own algorithm, \code{CloudTracker}).   

We compare the $\epsff$ within simulated GMCs against the predictions of several theories for turbulence-regulated star formation.   None of these reproduces the magnitude of the surge in $\epsff$ that accompanies the end of feedback. For example, the KM, HC and PN models predict an increase by a factor of 2-3, but measured values from the simulations indicate an increase by more than an order of magnitude. Although our resolution of individual clouds is relatively poor, the surge in $\epsff$ exists after averaging all clouds, and is just as clear for our most massive clouds, and for the galaxy as a whole. 

Our experiment reveals that, apart from its role as a regulator of star formation, stellar feedback affects almost all aspects of the population and internal states of GMCs.  A galaxy that loses feedback quickly sprouts a population of new clouds, especially bound GMCs with low virial parameters, and its unbound clouds show a marked drop in their internal turbulence.  GMCs become denser overall, although neither this nor the increase in their number is enough to explain the star formation surge.   
Stellar feedback also destroys low and intermediate mass clouds, but we find that it is not the dominant mechanism in our cloud sample. We suspect this has to do with how we classify mergers between unequal mass clouds as the death of the lower mass cloud. Taken together, these points imply that stellar feedback is an essential ingredient of cloud formation and destruction, and of the forces that regulate star formation within clouds.  

While our results remain to be verified with better-resolved simulations, they strongly suggest that the turbulence-regulated star formation theories that we assess in \autoref{S:Theory-Comp}  are lacking something important. Stellar feedback is one obvious missing ingredient -- indeed, it is the essential ingredient of feedback-regulated star formation theories. We argue that on scales of entire clouds or larger, ``feedback-aware'' models such as those of \citet{ostriker2011}, \citet{kim2011}, \citet{orr2019}, and \citet{ostriker2022} may be more applicable for predicting star formation efficiencies. There are ways to overcome the issues in turbulence-regulated models, by incorporating time-dependence \citep[e.g.][etc.]{guszejnov2015a, guszejnov2016, guszejnov2016a, Jaupart2020} or a more accurate functional form of the density PDF \citep{BB17, Burkhart2018, Burkhart2019} for example.  Further development will be required to permit a clear comparison with simulations such as ours.

\section*{Acknowledgements}

We thank Eve Ostriker, Enrique Vazquez-Semadeni, Vadim Semenov, Marta Reina-Campos, Matt Orr, James Beattie and Raghav Arora for useful conversations. We also thank Christoph Federrath and the anonymous referee for comments which helped to improve the manuscript. The research of SK, CM, and NM was supported by Discovery Grants from the National Science and Engineering Research Council.  NM was supported by a Canada Research Chair (Tier 1). Computations were performed on the Niagara supercomputer at the SciNet HPC Consortium. SciNet is funded by Innovation, Science and Economic Development Canada; the Digital Research Alliance of Canada; the Ontario Research Fund: Research Excellence; and the University of Toronto. This work was performed in part at the Aspen Center for Physics, which is supported by National Science Foundation grant PHY-2210452 Support for MYG was provided by NASA through the NASA Hubble Fellowship grant \#HST-HF2-51479 awarded  by  the  Space  Telescope  Science  Institute,  which  is  operated  by  the   Association  of  Universities  for  Research  in  Astronomy,  Inc.,  for  NASA,  under  contract NAS5-26555. The data used in this work were, in part, hosted on facilities supported by the Scientific Computing Core at the Flatiron Institute, a division of the Simons Foundation. AW received support from: NSF via CAREER award AST-2045928 and grant AST-210772: NASA ATP grant 80NSSC20K0513: HST grants AR-15809, GO-15902, GO-16273 from STScI.

\vspace{5mm}







\bibliographystyle{aasjournal}
\bibliography{main}{}



\end{document}